\documentclass[smallcondensed]{svjour3}     % onecolumn (ditto)
\smartqed  % flush right qed marks, e.g. at end of proof
\usepackage{graphicx}
\usepackage[latin1]{inputenc}
\usepackage[T1]{fontenc}
\usepackage[english]{babel}
\usepackage{amsfonts}
\usepackage{amssymb}
\usepackage{dsfont}
\usepackage{hyperref}
\usepackage{amsmath}
\usepackage{url}
\usepackage{appendix}
\usepackage{natbib}

\newcommand{\E}{\mathds{E}}

\renewcommand{\(}{\left(}
\renewcommand{\)}{\right)}

\newcommand{\1}{\mathds{1}}

\renewcommand{\geq}{\geqslant}
\renewcommand{\leq}{\leqslant}

\begin{document}

\title{Stochastic areas of diffusions and applications in risk theory
}
%\subtitle{}

\titlerunning{Stochastic area of diffusion}        % if too long for running head

\author{ Zhenyu Cui  }

%\institute{Zhenyu Cui \at
%              department of Mathematics at Brooklyn College of the City University of New York \\
%              Tel.: +1718-951-5600, ext. 6892\\
%              Fax: +1718-951-4674\\
%              \email{zhenyucui@brooklyn.cuny.edu}           %  \\
%
%}

\institute{Z. Cui
           \at Corresponding Author. Department of Mathematics, \\
              Brooklyn College of the City University of New York, \\
              Tel.: +1718-951-5600, ext. 6892\\
              Fax: +1718-951-4674\\
              \email{zhenyucui@brooklyn.cuny.edu}
}

\date{Received: date / Accepted: date}

\maketitle

\begin{abstract}
In this paper we study the stochastic area swept by a regular time-homogeneous diffusion till a stopping time. This unifies some recent literature in this area. Through stochastic time change we establish a link between the stochastic area and the stopping time of another associated time-homogeneous diffusion. Then we characterize the Laplace transform of the stochastic area in terms of the eigenfunctions of the associated diffusion. We also explicitly obtain the integer moments of the stochastic area in terms of scale and speed densities of the associated diffusion. Specifically we study in detail three stopping times: the first passage time to a constant level,  the first drawdown time and the Azema-Yor stopping time. We also study the total occupation area of the diffusion below a constant level.  We show applications of the results to a new structural model of default (Yildirim \citeyear{Y06}), the Omega risk model of bankruptcy in risk analysis (Gerber, Shiu and Yang \citeyear{GSY12}), and a diffusion risk model with surplus-dependent tax (Albrecher and Hipp \citeyear{AH07}, Li, Tang and Zhou \citeyear{LTZ13}).
\end{abstract}

\vspace{3mm}



\vspace{3mm}

\maketitle

\newpage

\section{Introduction}\label{intro}
The stochastic area swept by a stochastic process till a stopping time has an intuitive geometric meaning and has been applied to modeling objects of ``accumulative nature" in physics, queueing theory, mathematical finance, actuarial risk theory and graph enumeration problems. Let $(V_t)_{\left\{t\geq 0\right\}}$ denote the underlying stochastic process.

 %We also consider the stochastic area till an Azema-Yor stopping time.

In queueing theory, if $V_t$ represents the length of a queue at time $t$, and denote the  first passage time of $V$ to the threshold $l$ by $\tau_l$, which is the time when the queue system first collapses(\textit{overflow time}). Then the stochastic area swept by $V$ until $\tau_l$ represents the cumulative waiting time experienced by all the users till the overflow of the system.

In mathematical finance, there are two main approaches in the area of credit risk modeling. The structural approach(Merton \citeyear{M74}, Black and Cox \citeyear{BC76}) assumes that the default event happens when the total value of the firm's asset first goes below the face value of its debt. If we model the firm's asset value as a diffusion $V$, then the default time is the first passage time of $V$ to a fixed threshold. The second is the reduced-form model(Jarrow and Turnbull \citeyear{JT95}, Duffie and Singleton \citeyear{DS}), where default is modeled by an exogenously-determined intensity or compensator process. The default time is the first jump time of a point process. Recently there is a third new approach by
Yildirim \citeyear{Y06}, which uses the information of the stochastic area swept by the firm's asset value process.  The default time is when the cumulative stochastic area below a threshold level exceeds an exogenous level. This model with stochastic area allows the firm to recover from financial distress and is flexible.

In actuarial risk theory or ruin theory, recently there is some interest in using the total occupation time of a stochastic process below a constant level to model bankruptcy. The new model is named the ``Omega risk model". The Omega risk model was first introduced by Albrecher, Gerber and Shiu \citeyear{AGS11}, and it distinguishes the ruin time(negative surplus) from the bankruptcy time(negative surplus area surpasses a specified level) of a company. In classical ruin theory, the ruin event occurs when the surplus of a company first becomes negative. The Omega risk model assumes the more realistic situation that a company can still do business even with a negative surplus. The bankruptcy time is linked to the total occupation time of the firm value $V$ below a certain threshold $l$. Here the total occupation time $\int_0^{\infty}\1_{\left\{V_t <l \right\}}dt$ is a special stochastic area with the integrand being an indicator function.

In the applied probability literature, there are a few papers proposing to study the distribution of the area swept by a stochastic process till a stopping time. More specifically, Perman and Wellner \citeyear{PW96} study the distribution and moments of the stochastic area in the case of a standard Brownian motion. Kearney and Majumadar \citeyear{KM05}, Kearney, Majumadar and Martin \citeyear{KMM07} extend the study to a drifted Brownian motion. Knight \citeyear{K2000} considers the case of a reflected Brownian bridge. Janson \citeyear{J07} provides a survey to this area and links it to the Wright's constants in graph enumeration. Recently, Abundo \citeyear{A13} considers a one-dimensional jump diffusion and show that the Laplace transform and moments of the associated stochastic area till the first passage time to zero are solutions to certain partial differential-difference equations(PDDE) with outer conditions. Abundo \citeyear{A133} extends similar results to the stochastic area till the first passage time to a non-zero level.

Starting from the seminal paper of Lehoczky \citeyear{L77}, there is a continued interest in the study of drawdown/drawup time of diffusion processes(Hadjiliadis and Vecer \citeyear{HV06}, Pospisil, Vecer and Hadjiliadis \citeyear{PVH09}, Zhang and Hadjiliadis (\citeyear{ZH10}, \citeyear{ZH12})). For a pointer to the vast literature in this area, please refer to the Ph.D. thesis of Zhang \citeyear{Z10}. The stochastic area till the first drawdown time represents the accumulated firm value till the drawdown event, and reflects the profitability and solvency of a firm during a financial crisis. The Azema-Yor stopping time is a generalized drawdown time and is a candidate solution to many optimal stopping problems(Graversen and Peskir \citeyear{GP97}, Pedersen \citeyear{P05}, Shepp and Shiryaev \citeyear{SS93}), and also to the Skorokhod embedding problem(Obloj \citeyear{O04}). There is some recent interest in the literature to study risk models with tax: risk models perturbed by a functional of the running maximum process.  Albrecher and Hipp \citeyear{AH07} introduce a constant tax rate at profitable times of the compound Poisson risk model. Li, Tang and Zhou \citeyear{LTZ13} introduce a surplus-dependent tax rate to a diffusion risk model and model the ruin time of the company by the two-sided exit time of the diffusion process. Refer to Kyprianou and Ott \citeyear{KO12} for a pointer to relevant literature.

We make three contributions to the current literature. First, through stochastic time change we link the study of the stochastic area till a stopping time to the study of a related stopping time of another time-homogeneous diffusion, whose drift and diffusion coefficients can be explicitly determined. This provides a unified method to study the  stochastic area and allows us to characterize the Laplace transform of the stochastic area in terms of eigenfunctions of diffusions. In some examples, we can explicitly compute the eigenfunctions and the Laplace transform. We provide an explicit closed-form formula for the integer moments of the first passage area and extend the approach in Abundo \citeyear{A13}, \citeyear{A133}, since we do not need to solve the associated partial differential-difference equations, but solve a Sturm-Liouville type ordinary differential equation(ODE) instead. There are a number of known examples where we can solve the ODE explicitly(see Borodin and Salminen \citeyear{BS02}). Second, we compute the probability of bankruptcy and the expected  time of ruin in the Omega risk model for diffusions with a general bankruptcy rate function $\omega(.)$. Previous literature considers the special case when $\omega(.)$ is a constant or a piecewise constant(Albrecher, Gerber and Shiu \citeyear{AGS11}, Gerber, Shiu and Yang \citeyear{GSY12}, and Li and Zhou \citeyear{LZ13}). We extend the literature to a more general bankruptcy rate function and this provides more flexibility in modeling. Third, we explicitly compute the expected time of ruin in a diffusion risk model with surplus-dependent tax rate.  We also explicitly characterize the expected stochastic area till the Azema-Yor stopping time, which represents the accumulated firm value till the optimal execution time in some optimal stopping problems(e.g. Russian options).

The paper is organized as follows: Section \ref{s2} presents the stochastic time change method and links the study of the stochastic area till a stopping time to the study of a related stopping time of another diffusion. Section \ref{s3} considers the stochastic area till the first passage time of the diffusion to a constant level. We compute the Laplace transform of the stochastic area in terms of eigenfunctions of the associated diffusions and express all its integer moments in terms of the scale and speed densities as an explicit recursion formula. We illustrate an example in the new structural model of default proposed in Yildirim \citeyear{Y06}. Section \ref{s4} provides the Laplace transform of the stochastic occupation area, and apply it to calculating the probability of bankruptcy in the Omega risk model with a general bankruptcy rate function $\omega(.)$. We illustrate with a concrete example. Section \ref{s5} computes the Laplace transform of the stochastic area till the first drawdown/drawup time of a diffusion. We obtain explicit closed-form formula for the expected time of ruin in a diffusion risk model with surplus-dependent tax.  Section \ref{s6} concludes the paper with future research directions.

\section{Stochastic time change and stochastic area till a stopping time }\label{s2}

Given a complete filtered probability space $(\Omega, \mathcal{F}, \mathcal{F}_t, P)$ with state space $J=(l,r), -\infty\leq l<r\leq \infty$,  and assume that the $J$-valued diffusion $V=(V_t)_{\left\{t\in[0,\infty)\right\}}$ satisfies the SDE
\begin{equation}
dV_t=\mu(V_t)\, dt\, +\, \sigma(V_t)\,  dW_t, \quad V_0=v_0\in J. \label{G}
\end{equation}
where $W$ is a $\mathcal{F}_t$-Brownian motion and $\mu, \sigma: J\rightarrow \mathbb{R}$ are Borel functions satisfying the Engelbert-Schmidt conditions
\begin{align}
\forall x\in J, \text{ } \sigma(x)\neq 0,\quad  \frac{1}{\sigma^2(\cdot)}, \quad \frac{\mu(\cdot)}{\sigma^2(\cdot)}\in L_{loc}^1 (J),\label{cond1}
\end{align}
where $L_{loc}^{1} (J)$ denotes the class of locally integrable functions, i.e. the functions $J\rightarrow \mathbb{R}$ are integrable on compact subsets of $J$. This condition \eqref{cond1} guarantees that the SDE \eqref{G} has a unique in law weak solution that possibly exits its state space $J$(see Theorem $5.15$, p$341$, Karatzas and Shreve \citeyear{KS91}).

 Denote the possible explosion time of $V$ from its state space by $\zeta$, i.e. $\zeta=\inf\{u>0, V_u \not\in J\} $,
  which means that $P$-a.s. on $\{\zeta=\infty\}$ the trajectories of $V$ do not exit $J$, and $P$-a.s. on $\{\zeta<\infty\}$,
  we have $\lim\limits_{t\rightarrow \zeta}V_t =r$ or $\lim\limits_{t\rightarrow \zeta}V_t =l$. $V$ is defined such that it stays at its
  exit point, which means that $l$ and $r$ are absorbing boundaries. The following terminology is used: $V$ exits the state space $J$ at $r$
  means $P(\zeta<\infty, \lim\limits_{t\rightarrow \zeta}V_t=r)>0$.

In what follows, $\lambda(.)$ denotes the Lebesgue measure on $B(\mathbb{R})$. Let $b$ be a Borel function such that $\lambda (x\in (l, r): b^2(x) =0)=0$, and assume the following local integrability condition
\begin{align}
 \forall x\in J, \text{ } \sigma(x)\neq 0,\quad  \frac{b^2(\cdot)}{\sigma^2(\cdot)} \in L_{loc}^1 (J).\label{cond2}
\end{align}
\begin{lemma}\label{ct}
Define the function  $\varphi_t:=\int_0^t b^2(V_u)du, \text{ for }t\in [0,\zeta]$. Then $\varphi_t$ is a non-decreasing and continuously differentiable function for $t\in [0,\zeta]$ with positive derivative $P$-a.s.
\end{lemma}
\proof
Recall that $b(.)$ is a positive Borel function, thus $\varphi_t$ is an
 increasing function for $0\leq t\leq \zeta$. For $t\in [0,\zeta)$, it is clear that $\varphi_t$ is a continuous function, and the continuity of $\varphi_t$ at $t=\zeta$ on the set $\left\{\zeta<\infty\right\}$ follows from the Dambis-Dubins-Schwartz theorem(see the proof of Theorem $4.6$, p175, Karatzas and Shreve \citeyear{KS91}). Also note that $\varphi_t$ is represented as a time integral and is thus differentiable with derivative $b^2(V_t)$, which is positive $P$-a.s. from the definition of the function $b(.)$. \qed
%\end{proof}
 The following result is about stochastic time-change, and for completeness we provide its proof.
\begin{theorem}\label{tc}(Theorem $3.2.1$ of Cui \citeyear{C13})

Assume the conditions \eqref{cond1} and \eqref{cond2}

(i) Define
 \begin{align}
 \tau(t)&:=\tau_t :=
 \begin{cases}
 \inf\{u\geq 0 : \varphi_{u\wedge \zeta} >t\}, &\mbox{ on } \left\{0\leq t <\varphi_{\zeta}\right\},\\
 \infty, &\mbox{ on } \left\{\varphi_{\zeta}\leq t<\infty\right\}.\label{tt}
 \end{cases}
 \end{align}
 Define a new filtration $\mathcal{G}_{t}=\mathcal{F}_{\tau_t}, t\in[0,\infty)$, and a new $\mathcal{G}_t$-adapted process $X_t:=V_{\tau_t}$, on $\left\{0\leq t <\varphi_{\zeta}\right\}$. Then  we have the stochastic representation
\begin{align}
V_{t} &=X_{\int_{0}^{t} b^2(V_{s})ds}=X_{\varphi_{t}}, \quad P-a.s., \quad \text{ on } \left\{0\leq t <\zeta\right\}.\label{tc1}
\end{align}
and the process $X$ is a time-homogeneous diffusion, which solves the following SDE under $P$
\begin{align}
dX_{t} &=\frac{\mu(X_t)}{b^2(X_t)}\1_{\left\{t\in [0,\varphi_\zeta)\right\}}dt +\frac{\sigma(X_t)}{b(X_t)}\1_{\left\{t\in [0,\varphi_\zeta)\right\}}dB_t, \quad X_0=v_0.\label{xsde2}
\end{align}
where $B_t$ is the $\mathcal{G}_t$-adapted Dambis-Dubins-Schwartz Brownian motion defined in the proof.

(ii) Define $\zeta^X :=\inf\left\{u>0: X_u \not \in J \right\}$, then $\zeta^X =\varphi_{\zeta}=\int_0^{\zeta} m^2 (V_s )ds$, $P$-a.s., and we can rewrite the SDE \eqref{xsde2}  as
\begin{align}
dX_{t} &=\frac{\mu(X_t)}{b^2(X_t)}\1_{\left\{t\in [0,\zeta^X)\right\}}dt +\frac{\sigma(X_t)}{b(X_t)}\1_{\left\{t\in [0,\zeta^X)\right\}}dB_t, \quad X_0=v_0.\label{xsde}
\end{align}

(iii) Let $\tau$ denote a $\mathcal{F}_t$ stopping time\footnote{It is understood that $\tau=\zeta$, $P$-a.s on $\left\{ \tau>\zeta \right\}$.} of $V_t, t\in[0,\zeta)$, then $\varphi_{\tau}:=\int_0^{\tau} b^2(V_s)ds$ is a $\mathcal{G}_t$ stopping time and $\tau^X=\varphi_{\tau}$, P-a.s., where $\tau^X$ is the corresponding stopping time for $X_t, t\in[0,\zeta^X)$.
\end{theorem}
\proof
Since $\lambda  (x\in (\ell,r): b^2(x) =0)=0$, $\varphi_s$ is an increasing and continuous function on $[0,\zeta]$, from Problem $3.4.5$ (ii), p174 of Karatzas and Shreve \citeyear{KS91},  $\varphi_{\tau_t \wedge \zeta}=t\wedge \varphi_{\zeta}$, $P$-a.s. for $0\leq t<\infty$. On $\left\{0\leq t<\varphi_{\zeta}\right\}$, when $u=\zeta$,   $\varphi_{\zeta\wedge \zeta}=\varphi_{\zeta}>t$ holds $P$-a.s. according to the assumption. Then  $\tau_t\leq \zeta$, $P$-a.s. due to the definition in \eqref{tt}. Thus $\varphi_{\tau_t}=t$, $P$-a.s. on $\left\{0\leq t<\varphi_{\zeta}\right\}$.

On $\left\{0\leq s<\zeta\right\}$, choose $t=\varphi_s$, then $0\leq t <\varphi_{\zeta}$, $P$-a.s. Substituting this $t$ into the definition of the process $X$,  $X_{\varphi_s}=X_t:=V_{\tau_t}=V_{\tau_{\varphi_s}}=V_s$, $P$-a.s. For the last equality, note that $\tau_{\varphi_s}=\inf\{u\geq 0:\varphi_{u\wedge\zeta}>\varphi_s \}=\inf\{u\geq 0:u\wedge\zeta>s \}=s$, $P$-a.s., on $\left\{0\leq s<\zeta\right\}$. Then we have proved the representation $V_s =X_{\varphi_s}$, on $\left\{0\leq s<\zeta\right\}$.

 For $X$ satisfying the relation \eqref{tc1}, we aim to show that $X$ satisfies the SDE \eqref{xsde2},
where $B$ is the Dambis-Dubins-Schwartz Brownian motion adapted to $\mathcal{G}_t$ constructed as follows:
Note that $M_{t\wedge \zeta}=\int_0^{t\wedge \zeta} b(V_u) dW_u, t\in[0,\infty)$ is a continuous local martingale, with quadratic variation
$\varphi_{t\wedge \zeta}=\int_0^{t\wedge \zeta} b^2(V_u) du, t\in[0,\infty)$.
Then $\lim_{t\rightarrow \infty}\varphi_{t\wedge \zeta}=\varphi_{\zeta}$, $P$-a.s. due to the left continuity of $\varphi_s$ at $s=\zeta$.
From the Dambis-Dubins-Schwartz theorem(Ch.V, Theorem $1.6$ and Theorem $1.7$ of Revuz and Yor \citeyear{RY99}), there exists
a possible enlargement $(\bar{\Omega},\bar{\mathcal{G}_t}, \bar{P})$ of $(\Omega, \mathcal{G}_{t}, P)$ and a standard Brownian motion $\bar{\beta}$ on $\bar{\Omega}$ independent of
$M$ with $\bar{\beta}_0=0$, such that the process
\begin{align}
B_{t} &:=
\begin{cases}
\int_0^{\tau_t} b(V_u) dW_u, &\mbox{ on } \left\{t < \varphi_{\zeta}\right\}, \\
\int_0^{\zeta} b(V_u) dW_u+\widetilde{\beta}_{t-\varphi_{\zeta}}, &\mbox{ on } \left\{t \geq \varphi_{\zeta}\right\}.\label{conb}
\end{cases}
\end{align}
is a standard linear $\mathcal{G}_t$-Brownian motion. Our construction of $\tau_t, t\in[0,\infty)$ agrees with that in Problem $3.4.5$, p174 of Karatzas and Shreve \citeyear{KS91}. From Problem $3.4.5$ (ii) and the construction \eqref{conb},  $B_{\varphi_s}=M_s$, $P$-a.s. on $\left\{0\leq s<\zeta\right\}$. On $\left\{s=\zeta\right\}$, $B_{\varphi_{\zeta}}:=\int_0^{\zeta} b(V_u) dW_u+\widetilde{\beta}_0=\int_0^{\zeta} b(V_u) dW_u=:M_{\zeta}$, $P$-a.s. Thus $B_{\varphi_t}=M_t$, $P$-a.s. on $\left\{0\leq t\leq \zeta\right\}$.

For the convenience of exposition, denote $\mu_1(.)=\mu(.)/b^2(.)$, and $\sigma_1(.)=\sigma(.)/b(.)$.
Integrate the SDE in \eqref{G} from $0$ to $t\wedge \zeta$
 \begin{align}
 V_{t\wedge \zeta} -V_0&=\int_0^{t\wedge \zeta} \mu(V_u) du +\int_0^{t\wedge \zeta} \sigma(V_u)dW_u\notag\\
 &= \int_0^{t\wedge \zeta} \mu_1(V_u)b^2 (V_u)du +\int_0^{t\wedge \zeta} \sigma_1(V_u)b(V_u)dW_u.\label{intermet2f}
 \end{align}

Apply the change of variables formula similar to Problem $3.4.5$ (vi), p174 of Karatzas and Shreve \citeyear{KS91}, and note the relation \eqref{tc1}
\begin{align}
\int_0^{t\wedge \zeta} \mu_1(V_{u})b^2(V_{u})du&= \int_0^{t\wedge \zeta} \mu_1(X_{\varphi_{u}})d\varphi_{u}=\int_0^{\varphi_{t\wedge \zeta}} \mu_1(X_u)du, \label{pu1f}
\end{align}
and similarly
\begin{align}
\int_0^{t\wedge \zeta} \sigma_1(V_u)b(V_{u})dW_u&=\int_0^{t\wedge \zeta} \sigma_1 (X_{\varphi_u}) dB_{\varphi_u}
= \int_0^{\varphi_{t\wedge \zeta}} \sigma_1 (X_u)dB_u \label{pu2f}
\end{align}
where the first equality in \eqref{pu2f} is due to the relationship $B_{\varphi_u}=M_u=\int_0^u b(V_s)dW_s$, $P$-a.s. on $\left\{0\leq u \leq t\wedge\zeta\right\}$, which we have established above.
Also notice the representation $V_{t\wedge \zeta}=X_{\varphi_{t\wedge \zeta}}$, $P$-a.s. and $V_0=X_0$, then
\begin{align}
X_{\varphi_{t\wedge \zeta}} -X_0 &=\int_0^{\varphi_{t\wedge \zeta}} \mu_1(X_u)du +\int_0^{\varphi_{t\wedge \zeta}} \sigma_1 (X_u)dB_u \label{intermetf}
\end{align}
Then on $\left\{0\leq s \leq \varphi_{t\wedge \zeta}\right\}$
\begin{align}
X_s -X_0 &=\int_0^s \mu_1(X_u)du +\int_0^s \sigma_1 (X_u)dB_u.\label{a3f}
\end{align}
Note that for $0\leq t<\infty$, we have $s\in[0, \varphi_{\zeta}]$, $P$-a.s. From \eqref{a3f}, and recall the definition of $\mu_1(.)$ and $\sigma_1(.)$, we have the following SDE for $X$
\begin{align}
dX_{s} &=\frac{\mu(X_s)}{b^2(X_s)}\1_{\left\{s\in [0,\varphi_\zeta)\right\}}ds +\frac{\sigma(X_s)}{b(X_s)}\1_{\left\{s\in [0,\varphi_\zeta)\right\}}dB_s,\quad  X_0=V_0=v_0. \notag
\end{align}
This completes the proof of statement (i).

Statement (ii) and (iii) are direct consequences of the stochastic representation $V_{t\wedge \zeta}=X_{\varphi_{t\wedge \zeta}}$, $P$-a.s. in statement (i), because $\varphi_t$ is an increasing function with respect to $t$. This completes the proof.\qed
%\end{proof}

\section{Stochastic first passage area and moments }\label{s3}

In this section, we consider the two-sided exit time of the diffusion in \eqref{G} from an open interval $(a,c)\subset \bar{J}$ such that $a<v_0<c$. Define
\begin{align}
\tau_x&=\inf \{t\geq 0: V_t =x\}, \quad x\in \bar{J},
\end{align}
where $\inf\emptyset =\infty$ by convention.

Now we recall some classical theory on diffusion exit times. Define the scale density of the diffusion $V$ in \eqref{G}
\begin{align}
s(x):=\exp\left\{ -\int_{.}^x \frac{2\mu(u)}{\sigma^2(u)}du \right\}, \quad x\in \bar{J},\label{scaleden}
\end{align}
and the scale function is
\begin{align}
S(x):=\int_{.}^x s(y) dy=\int_{.}^x \exp\left\{ -\int_{.}^y \frac{2\mu(u)}{\sigma^2(u)}du \right\}dy, \quad x\in \bar{J}.\label{scale}
\end{align}

The Laplace transforms of $\tau_a$ and $\tau_c$ of the two-sided exit problem for a diffusion process $V$ were first solved by Darling and Siegert \citeyear{DS53}. Consider the following Sturm-Liouville ordinary differential equation
\begin{align}
\frac{1}{2}\sigma^2(x) g^{\prime\prime}(x)+\mu(x)g^{\prime}(x)&=\lambda g(x), \quad \lambda\geq 0, \label{sl}
\end{align}
and from classical diffusion theory it can be shown that \eqref{sl} always has two independent, positive and convex solutions\footnote{See Borodin and Salminen \citeyear{BS02} for a collection of explicit examples of diffusions.}. Here we denote the decreasing solution as $g_{-,\lambda}(.)$, and the increasing solution as $g_{+,\lambda}(.)$. Based on this pair of solutions, define the auxiliary function
\begin{align}
f_{\lambda}(y,z)&=g_{-,\lambda}(y) g_{+,\lambda}(z)-g_{-,\lambda}(z)g_{+,\lambda}(y), %\quad w_{\lambda}(y,z)=\frac{\partial}{\partial z}f_{\lambda}(y,z).
\end{align}
We have the following lemma.
\begin{lemma}\label{DS}
(Theorem $3.2$ of Darling and Siegert \citeyear{DS53})

With $a<v_0<c$, and $\lambda\geq 0$, we have the following Laplace transforms
\begin{align}
\E_{v_0}[e^{-\lambda \tau_a}; \tau_a <\tau_c]&=\frac{f_{\lambda}(v_0,c)}{f_{\lambda}(a,c)},
\end{align}
and
\begin{align}
\E_{v_0}[e^{-\lambda \tau_c}; \tau_c <\tau_a]&=\frac{f_{\lambda}(a,v_0)}{f_{\lambda}(a,c)}.
\end{align}
\end{lemma}

The following result gives the Laplace transform of the stochastic area till the first passage time.
\begin{proposition}\label{sfpt}
With $a<v_0<c$, and $\lambda\geq 0$, we have the following Laplace transforms
\begin{align}
\E_{v_0}[e^{-\lambda \int_0^{\tau_a} b^2(V_s)ds}; \tau_a <\tau_c]&=\frac{f^{\ast}_{\lambda}(v_0,c)}{f^{\ast}_{\lambda}(a,c)},\notag\\
\E_{v_0}[e^{-\lambda \int_0^{\tau_c} b^2(V_s)ds}; \tau_c <\tau_a]&=\frac{f^{\ast}_{\lambda}(a,v_0)}{f^{\ast}_{\lambda}(a,c)},
\end{align}
and
\begin{align}
f^{\ast}_{\lambda}(y,z)&=g^{\ast}_{-,\lambda}(y) g^{\ast}_{+,\lambda}(z)-g^{\ast}_{-,\lambda}(z)g^{\ast}_{+,\lambda}(y),
\end{align}
where $g^{\ast}_{-,\lambda}(.)$ and  $g^{\ast}_{+,\lambda}(.)$ are respectively the decreasing and increasing solutions of the following Sturm-Liouville type ordinary differential equation
\begin{align}
\frac{1}{2}\frac{\sigma^2(x)}{b^2(x) } g^{\prime\prime}(x)+\frac{\mu(x)}{b^2(x)} g^{\prime}(x)&=\lambda g(x), \quad \lambda\geq 0,\label{sl2}
\end{align}
\end{proposition}
\begin{proof}
From Theorem \ref{tc} (i), we have $V_t = X_{\int_0^t b^2(V_s) ds}$, $P$-a.s. on $\left\{0\leq t <\zeta  \right\}$. Define
\begin{align}
\tau^{X}_y&=\inf \{t\geq 0: X_t =y\},
\end{align}
then from Theorem \ref{tc} (iii), we have $\tau^{X}_y=\int_0^{\tau_y} b^2(V_s)ds$, $P$-a.s. Also notice the equivalence of events: $\left\{\tau_a <\tau_c\right\}$ and $\left\{\tau^X_a <\tau^X_c\right\}$.  Then we have
\begin{align}
\E_{v_0}\left[e^{-\lambda \int_0^{\tau_a} b^2(V_s)ds}; \tau_a <\tau_c\right]&=\E_{v_0}[e^{-\lambda \tau^X_a}; \tau^X_a <\tau^X_c];\notag\\
\E_{v_0}[e^{-\lambda \int_0^{\tau_c} b^2(V_s)ds}; \tau_c <\tau_a]&=\E_{v_0}[e^{-\lambda \tau^X_c }; \tau^X_c <\tau^X_a],\label{key1}
\end{align}
and we have transformed the study of the stochastic area till the first passage time to a related problem of the first passage time of the diffusion $X$. Note that $X$ is also a time-homogeneous diffusion with SDE given in \eqref{xsde}. Then \eqref{key1} combined with Lemma \ref{DS} completes the proof.
\qed
\end{proof}

\begin{remark}
Theorem $2.3$ of Abundo \citeyear{A13} provides the partial differential-difference equation(PDDE) that the Laplace transform of the stochastic area should satisfy. In the case of diffusions with no jumps, the formulation in $(3.4)$ of Abundo \citeyear{A13} includes outer conditions. Our approach is directly based on the classical diffusion theory developed in Darling and Siegert \citeyear{DS53}, and the Laplace transform is expressed using the eigenfunctions of diffusions.  He considers the case of the first passage time to the level $0$, and we generalize it to a possibly non-zero constant here.  Note that Abundo \citeyear{A133} considers the case of first passage time to a possibly non-zero constant level, but the method still involves solving an associated PDDE with outer conditions(see Theorem $2.3$, p5 of Abundo \citeyear{A133}).
\end{remark}

%\subsection{Moments of the stochastic area till the first passage time}
In subsequent discussion, consider the two-sided exit time $\tau=\tau_a \wedge \tau_c$, and the stochastic area $A_{\tau}:=\int_0^{\tau} b^2(V_s)ds$.
We aim to link the integer moments of the stochastic area to the moments of the two-sided exit time $\tau$ of the diffusion $X$ with SDE in \eqref{xsde}. Define the speed density of the diffusion $V$ in \eqref{G} as
\begin{align}
m(x)&=\frac{2}{\sigma^2(x) s(x)}, \quad x\in \bar{J},
\end{align}
where $s(.)$ is the scale density defined in \eqref{scaleden}.
First recall the following lemma using our notation.
\begin{lemma}\label{mo}
(Corollary 2.1, Wang and Yin \citeyear{WY08})

Define $\mu_n (x)=\E_{x}[ \tau^n ]$, then there is the following recursive relation
\begin{align}
\mu_n(x)&= n \frac{S(x)-S(a)}{S(c)-S(a)}\int_{x}^c (S(c)-S(y)) \mu_{n-1}(y) m(y) dy\notag\\
&+ n \frac{S(c)-S(x)  }{S(c)-S(a)} \int_a^{x} (S(y)-S(a))\mu_{n-1}(y)m(y)dy, \quad n=1,2,...\label{moment}
\end{align}
and
\begin{align}
\E_{x}[\tau]&=\frac{S(x)-S(a)}{S(c)-S(a)}\int_{x}^c (S(c)-S(y)) m(y) dy+ \frac{S(c)-S(x)}{S(c)-S(a)} \int_a^{x} (S(y)-S(a))m(y)dy.
\end{align}
\end{lemma}
\begin{remark}
Letting $a\rightarrow -\infty$ in \eqref{moment}, one obtains the Siegert's recursive formula\footnote{equation (3.14) of Siegert \citeyear{S51}} for the moments of the first passage time of $V_t$ through $c$.
\end{remark}
We have the following result for the integer moments of the stochastic area.
\begin{proposition}\label{mfpt}
Define $\mu^{\ast}_n (x)=\E_{x}[ \(\int_0^\tau b^2(V_s)ds \)^n ]=\E_{x}[ (A_{\tau})^n ]$, then there is the following recursive relation
\begin{align}
\mu^{\ast}_n(x)&= n \frac{S(x)-S(a)}{S(c)-S(a)}\int_{x}^c (S(c)-S(y)) \mu^{\ast}_{n-1}(y) m^{\ast}(y) dy\notag\\
&+ n \frac{S(c)-S(x)  }{S(c)-S(a)} \int_a^{x} (S(y)-S(a))\mu^{\ast}_{n-1}(y)m^{\ast}(y)dy, \quad n=1,2,...\label{moment2}
\end{align}
and
\begin{align}
\E_{x}[A_{\tau}]&=\frac{S(x)-S(a)}{S(c)-S(a)}\int_{x}^c (S(c)-S(y)) m^{\ast}(y) dy+ \frac{S(c)-S(x)}{S(c)-S(a)} \int_a^{x} (S(y)-S(a))m^{\ast}(y)dy,\notag
\end{align}
where
\begin{align}
m^{\ast}(x)&=\frac{2 b^2(x)}{\sigma^2(x) s(x)}, \quad x\in \bar{J},\label{sp2}
\end{align}
\end{proposition}
\begin{proof}
From Theorem \ref{tc} (i), we have $V_t = X_{\int_0^t b^2(V_s) ds}$, $P$-a.s. on $\left\{0\leq t <\zeta  \right\}$. Define $\tau^{X}=\tau^X_a \wedge \tau^X_b$, then from Theorem \ref{tc} (iii), we have $\tau^{X}=\int_0^{\tau} b^2(V_s)ds$, $P$-a.s. Note that $X$ and $V$ share the same scale density since $\frac{\mu(.)/b^2(.)}{\sigma^2(.)/b^2(.)}=\frac{\mu(.)}{\sigma^2(.)}$, but the speed density for diffusion $X$ is different and given as $m^{\ast}(x)$ in \eqref{sp2}.
This combined with Lemma \ref{mo} completes the proof.  \qed
\end{proof}
\begin{remark}
Abundo \citeyear{A13} derives the recursive ODEs for the moments of $A_{\tau}$ in equation (3.5), p94 of his paper. Here we have derived explicit recursive relations of all the integer moments in terms of scale and speed densities, and do not need to solve an ODE.
\end{remark}

\textbf{Example: case of the geometric Brownian motion in a new structural model of default}

The structural approach to credit risk modeling assumes that the default event happens when the total value of the firm's asset first goes below the face value of its debt. The reduced-form approach assumes that default is modeled by an exogenously-determined intensity or compensator process, and the default time is the first jump time of a point process. Recently there is a third new approach by
Yildirim \citeyear{Y06}, which uses the information of the stochastic area under the firm's asset value process.  The default time is when the following two events both happen: the firm value process hits an default level and the stochastic area till this hitting time exceeds an exogenous level. In order to determine the probability of default for this new structural model, we need to determine the distribution of the stochastic area till the first passage time.
If we model the firm's asset value as a geometric Brownian motion, then we aim to calculate the Laplace transform of the stochastic passage area.

Assume that $V_t, t\geq 0$ is a geometric Brownian motion with state space $J=(0,\infty)$
\begin{align}
dV_t&=\mu V_t dt +\sigma V_t dW_t, \quad V_0=v_0\in J,\label{gbms}
\end{align}
where $\mu\neq 0$. Choose $b^2(x)=x^2$, and the SDE governing the diffusion $X$ is
\begin{align}
dX_t &=\frac{\mu}{X_t}dt +\sigma dW_t, \quad X_0=v_0.\label{bess}
\end{align}
We recognize \eqref{bess} as the SDE of a (scaled) standard Bessel process, and $X_t=\sigma R^{(\nu)}_t, t\geq 0$ where $R^{(\nu)}$ is a standard Bessel process with index $\nu=\frac{2\mu}{\sigma^2}-1$. For convenience, we assume that $\frac{2\mu}{\sigma^2}> 1$, thus $\nu> 0$. From classical diffusion theory, the associated ODE \eqref{sl2} has two fundamental solutions(Proposition $6.2.3.1$, p$345$ of Jeanblanc, Yor and Chesney \citeyear{JYC09}):
\begin{align}
g^{\ast}_{+,\lambda}(x)&=c_1 I_{\frac{\nu-2}{2}} ( x\sqrt{2\lambda})x^{1-\frac{\nu}{2}};\quad g^{\ast}_{-,\lambda}(x)=c_2 K_{\frac{\nu-2}{2}} ( x\sqrt{2\lambda})x^{1-\frac{\nu}{2}}, \quad x\in \bar{J},
\end{align}
with two constants $c_1$ and $c_2$, where  $I(.)$ and $K(.)$ are respectively the modified Bessel functions of the first and second kinds. Compute the auxiliary functions
\begin{align}
s(x)&=c_3 x^{-\nu-1}, \quad  S(x)=-c_3 \frac{x^{-\nu}}{\nu}, \quad m^{\ast}(x)=\frac{1}{c_3}\frac{2x^{\nu+1}}{\sigma^2} \quad x\in\bar{J},
\end{align}
where $c_3$ is a constant. From Proposition \ref{sfpt}, we can compute
\begin{align}
\E_{v_0}[e^{-\lambda \int_0^{\tau_a} b^2(V_s)ds}; \tau_a <\tau_c]%&=\frac{f^{\ast}_{\lambda}(v_0,c)}{f^{\ast}_{\lambda}(a,c)}\notag\\
&=\(\frac{v_0}{a}\)^{1-\frac{\nu}{2}} \frac{K_{\frac{\nu-2}{2}} ( v_0\sqrt{2\lambda})I_{\frac{\nu-2}{2}} ( c\sqrt{2\lambda})-K_{\frac{\nu-2}{2}} ( c\sqrt{2\lambda})I_{\frac{\nu-2}{2}} ( v_0\sqrt{2\lambda}) }{K_{\frac{\nu-2}{2}} ( a\sqrt{2\lambda})I_{\frac{\nu-2}{2}} ( c\sqrt{2\lambda})-K_{\frac{\nu-2}{2}} ( c\sqrt{2\lambda})I_{\frac{\nu-2}{2}} ( a\sqrt{2\lambda})},\notag
\end{align}
and
\begin{align}
\E_{v_0}[e^{-\lambda \int_0^{\tau_c} b^2(V_s)ds}; \tau_c <\tau_a]%&=\frac{f^{\ast}_{\lambda}(a,v_0)}{f^{\ast}_{\lambda}(a,c)}\notag\\
&=\(\frac{v_0}{c}\)^{1-\frac{\nu}{2}} \frac{K_{\frac{\nu-2}{2}} ( a\sqrt{2\lambda})I_{\frac{\nu-2}{2}} ( v_0\sqrt{2\lambda})-K_{\frac{\nu-2}{2}} ( v_0\sqrt{2\lambda})I_{\frac{\nu-2}{2}} ( a\sqrt{2\lambda}) }{K_{\frac{\nu-2}{2}} ( a\sqrt{2\lambda})I_{\frac{\nu-2}{2}} ( c\sqrt{2\lambda})-K_{\frac{\nu-2}{2}} ( c\sqrt{2\lambda})I_{\frac{\nu-2}{2}} ( a\sqrt{2\lambda})}.\notag
\end{align}

The first moment of the stochastic area is as follows.
\begin{align}
\E_{v_0}\left[ \int_0^{\tau} b^2(V_s) ds  \right]%&=\frac{x^{-\nu}-a^{-\nu}}{c^{-\nu}-a^{-\nu}} \frac{(\nu+2)x^2 -\nu c^2 -2c^{-\nu}x^{\nu+2}}{\sigma^2(\nu+2)}+\frac{c^{-\nu}-x^{-\nu}}{c^{-\nu}-a^{-\nu}} \frac{(\nu+2)x^2 -\nu a^2 -2a^{-\nu}x^{\nu+2}}{\sigma^2(\nu+2)}\notag\\
&=\frac{\nu x^2(c^{-\nu}-a^{-\nu}) -\nu a^2 c^2 (c^{-\nu-2}-a^{-\nu-2})-\nu x^{-\nu} (c^{2}-a^{2})}{  (c^{-\nu}-a^{-\nu})  \sigma^2(\nu+2)},\notag
\end{align}
and the other higher order moments can be similarly obtained from Proposition \ref{mfpt}.

%\begin{align}
%\psi_{\lambda}^{+, \ast}(x)&= \frac{g^{\prime, \ast}_{+, \lambda}(x)}{g^{\ast}_{+, \lambda}(x)}\notag\\
%&=, \quad x\in \bar{J},
%\end{align}
%and
%\begin{align}
%\psi_{\lambda}^{-, \ast}(x)&=-\frac{g^{\prime, \ast}_{-, \lambda}(x)}{g^{\ast}_{-, \lambda}(x)}\notag\\
%&=, \quad x\in \bar{J},
%\end{align}
%and
%\begin{align}
%\psi_0^{-,\ast}(x)&= \frac{s(x)}{\int_x^{\infty} s(y) dy}\notag\\
%&=,
%\end{align}
%
%\begin{align}
%\psi_0^{+,\ast}(x)&= \frac{s(x)}{\int_{-\infty}^x s(y) dy}\notag\\
%&=,
%\end{align}
%Compute the scale function
%\begin{align}
%S(x)&=\frac{x^{\nu+2}}{\nu+2}, \quad x\in\bar{J}.
%\end{align}
%Since $\nu\geq 0$, it is clear that $S(\infty)=\infty$. Apply Proposition \ref{o1} and we have
%\begin{align}
%\E_{v_0} \left[ e^{-\int_0^{\infty}b^2(V_s)\1_{V_s<0}ds} \right]&=\E_{v_0} \left[ e^{-\sigma^2\int_0^{\infty}(R^{\nu}_s)^2\1_{R^{\nu}_s<0}ds} \right]\notag\\
%&=\frac{ \psi^{-, \ast}_{0}(0) }{\psi^{+, \ast}_{\sigma^2}(0)+\psi^{-, \ast}_{0}(0)}\notag\\
%&=
%\end{align}

\section{Stochastic occupation area and the Omega risk model }\label{s4}

Classical ruin theory assumes that the ruin will occur at the first time when the surplus of a company is negative. For a pointer to the literature in this area, please refer to Gerber and Shiu \citeyear{GS98}. Recently, a new concept of ruin has been proposed and studied in a series of papers starting with Albrecher, Gerber and Shiu \citeyear{AGS11}. They coined the name ``the Omega risk model", and within this model there is a distinction between \textit{ruin}(negative surplus) and \textit{bankruptcy}(going out of business). The company continues operation even with a period of negative surplus, and they introduce a bankruptcy rate function $\omega(x)$ with $x$ denoting the value of negative surplus. $\omega(.)$ can be treated as an intensity of bankruptcy, and for $x\leq 0$,  $\omega(x) dt$ is the probability of bankruptcy within $dt$ time units. Assume that the value of the company is modeled by a time-homogeneous diffusion $V_t, t\in[0,\zeta)$ with SDE \eqref{G}, and assume that the state space is $J=(l,r)$ with $-\infty\leq l<r\leq \infty$. Assume that the initial value of the company satisfies $v_0>0$. If we introduce an auxiliary ``bankruptcy monitoring" process $N$ on the same probability space(with a possibly enlarged filtration), and assume that conditional on $V$, $N$ follows a Poisson process with state-dependent intensity $\omega(V_t) \1_{\left\{V_t<0\right\}}, t>0$. Then we define the time of bankruptcy $\tau_{\omega}$ as the first arrival time of the Poisson process $N$, i.e.
\begin{align}
\tau_{\omega}:=\inf\left\{t\geq 0: \int_0^t \omega(V_s)\1_{\left\{V_s<0\right\}}ds >e_1  \right\},
\end{align}
where $e_1$ is an independent exponential random variable with unit rate. Define $e_{\lambda}$ is another independent exponential random variable with rate $\lambda$. We can express the Laplace transform of the bankruptcy time as
\begin{align}
\E_{v_0}[ e^{-\lambda \tau_{\omega}}]&=P_{v_0}(\tau_{\omega}<e_{\lambda})=1-\E_{v_0}\left[ e^{-\int_0^{e_{\lambda}} \omega(V_s)\1_{\left\{V_s<0\right\}}ds } \right].\label{lz3}
\end{align}
for $\lambda\geq 0$. Similar as in Gerber, Shiu and Yang \citeyear{GSY12}, define the (total) \textit{exposure} as
\begin{align}
\mathcal{E}:=\int_0^{\infty} \omega(V_s)\1_{\left\{V_s<0\right\}}ds.\label{ar}
\end{align}

Let $\lambda\rightarrow 0+$ in \eqref{lz3}, we can calculate the \textit{probability of bankruptcy} as
\begin{align}
\psi(v_0)&=P(\tau_{\omega}<\infty\mid V_0=v_0)=1-\E_{v_0}\left[e^{- \int_0^{\infty} \omega(V_s)\1_{\left\{V_s<0\right\}}ds }\right]=1-\E_{v_0}[e^{-\mathcal{E}}],\label{st}
\end{align}

%For an arbitrary bankruptcy function $\omega(x), x<0$, Gerber, Shiu and Yang \citeyear{GSY12} shows that $\psi(u)$ satisfies an ordinary differential equation given in equation $(40)$ of their paper
In the literature, people have considered the following special cases(see Sec. 4, Gerber, Shiu and Yang \citeyear{GSY12}): (i) $\omega(x)=c$ with a constant $c$ ; (ii)\footnote{The assumption of $\omega(x)$ being a piecewise constant is also employed in  Li and Zhou \citeyear{LZ13}. } $\omega(x)=\eta_k,$ if $c_{k-1}<x<c_k, k=1,2,...,n$ for the constants $c_0=-\infty <c_1<...<c_{n-1}<c_n=0$; (iii)\footnote{In this case, Gerber, Shiu and Yang \citeyear{GSY12} manage to express the probability of bankruptcy in terms of \textit{Airy functions}, but this is possible only in their setting of modeling firm value $V$ as an arithmetic Brownian motion with drift. } $\omega(x)=-\eta x, x<0$ for some $\eta>0$.

Intuitively, \eqref{ar} represents the ``stochastic occupation area", and it measures the area swept by the sample path of $V$ that lies under the level zero. This motivates us to study the general case and we assume an arbitrary bankruptcy rate function such that $\omega(x)\geq 0, x\leq 0$, $\omega(x)=0, x>0$ and $\omega(.)$ is a decreasing function. From \eqref{st}, to calculate the probability of bankruptcy, we aim to calculate the Laplace transform of the total exposure $\mathcal{E}$.

Based on the solutions to the ODE in \eqref{sl}, define a pair of Laplace exponents for $\lambda \geq 0$
\begin{align}
\psi_{\lambda}^{\pm}(x)&=\pm \frac{g^{\prime}_{\pm, \lambda}(x)}{g_{\pm, \lambda}(x)}, \quad x\in \bar{J},\label{le}
\end{align}
From properties of the solutions to the ODE \eqref{sl},  we have(see Li and Zhou \citeyear{LZ13})
\begin{align}
\psi_0^{-}(x)&= \frac{s(x)}{\int_x^{\infty} s(y) dy}, \quad \text{and} \quad \psi_0^{+}(x)= \frac{s(x)}{\int_{-\infty}^x s(y) dy}, \label{le2}
\end{align}

Recall the following result from Li and Zhou \citeyear{LZ13}.

\begin{lemma}
(Corollary $3.2$ of Li and Zhou \citeyear{LZ13})

For $v_0>0$
\begin{align}
\E_{v_0} \left[ e^{-\lambda\int_0^{e_{\delta}}\1_{\left\{V_s<0\right\}}ds} \right]&=\frac{g_{-,\delta}(v_0)}{g_{-,\delta}(0)}  \frac{ \frac{\delta}{\delta+\lambda} \psi^{+}_{\delta+\lambda}(0) +\psi^{-}_{\delta}(0) }{\psi^{+}_{\delta+\lambda}(0)+\psi^{-}_{\delta}(0)}+1-\frac{g_{-,\delta}(v_0)}{g_{-,\delta}(0)},\label{lz1}
\end{align}
and for $v_0\leq 0$,
\begin{align}
\E_{v_0} \left[ e^{-\lambda\int_0^{e_{\delta}}\1_{\left\{V_s<0\right\}}ds} \right]&=\frac{g_{+,\delta+\lambda}(v_0)}{g_{+,\delta+\lambda}(0)} \frac{ \frac{\delta}{\delta+\lambda} \psi^{+}_{\delta+\lambda}(0) +\psi^{-}_{\delta}(0) }{\psi^{+}_{\delta+\lambda}(0)+\psi^{-}_{\delta}(0)}+\frac{\delta}{\delta+\lambda}\(1-\frac{g_{+,\delta+\lambda}(v_0)}{g_{+,\delta+\lambda}(0)}\).\label{lz2}
\end{align}
\end{lemma}
Recall from Li and Zhou \citeyear{LZ13} the following property of the function $g_{-,\delta}(x)$ as $\delta\rightarrow 0+$.
\begin{align}
g_{-,0}(x)&=1, \quad \text{if } \int_x^{\infty}s(y)dy =\infty;\notag\\
g_{-,0}(x)&=\int_x^{\infty}s(y)dy, \quad \text{if } \int_x^{\infty}s(y)dy <\infty.
\end{align}
Now by taking $\delta\rightarrow 0+$ in the above \eqref{lz1} and \eqref{lz2}, we have

\begin{lemma}\label{w1}
For $v_0>0$, if\footnote{Since $s(y)>0$ on the compact interval $[0,v_0], v_0>0$ (or $[v_0,0], v_0<0$), $\int_{v_0}^{\infty}s(y)dy <\infty$ is equivalent to $S(\infty)=\int_{0}^{\infty}s(y)dy <\infty$. Similar for the other case.} $S(\infty) <\infty$, then
\begin{align}
\E_{v_0} \left[ e^{-\lambda\int_0^{\infty}\1_{\left\{V_s<0\right\}}ds} \right]&=1- \frac{\int_{v_0}^{\infty}s(y)dy}{\int_0^{\infty}s(y)dy} \frac{ \psi^{+}_{\lambda}(0) }{\psi^{+}_{\lambda}(0)+\psi^{-}_{0}(0)},\label{lz3}
\end{align}
and if $S(\infty) =\infty$, then
\begin{align}
\E_{v_0} \left[ e^{-\lambda\int_0^{\infty}\1_{\left\{V_s<0\right\}}ds} \right]&=\frac{ \psi^{-}_{0}(0) }{\psi^{+}_{\lambda}(0)+\psi^{-}_{0}(0)},\label{lz4}
\end{align}

For $v_0< 0$,
\begin{align}
\E_{v_0} \left[ e^{-\lambda\int_0^{\infty}\1_{\left\{V_s<0\right\}}ds} \right]&=\frac{g_{+,\lambda}(v_0)}{g_{+,\lambda}(0)} \frac{ \psi^{-}_{0}(0) }{\psi^{+}_{\lambda}(0)+\psi^{-}_{0}(0)}.\label{lz5}
\end{align}
\end{lemma}
For the diffusion $X$ with SDE in \eqref{xsde},  from the solutions to the associated Sturm-Liouville ODE in \eqref{sl2}, we define a pair of Laplace exponents for $\lambda \geq 0$
\begin{align}
\psi_{\lambda}^{\pm, \ast}(x)&=\pm \frac{g^{\prime, \ast}_{\pm, \lambda}(x)}{g^{\ast}_{\pm, \lambda}(x)}, \quad x\in \bar{J},\label{le3}
\end{align}
From properties of the solutions to the ODE \eqref{sl2},  we have
\begin{align}
\psi_0^{-, \ast}(x)&= \frac{s(x)}{\int_x^{\infty} s(y) dy}, \quad \text{and} \quad \psi_0^{+, \ast}(x)= \frac{s(x)}{\int_{-\infty}^x s(y) dy}, \label{le4}
\end{align}
and note that $\psi_0^{\pm, \ast}(x)=\psi_0^{\pm}(x)$ because the diffusions $V$ and $X$ share the same scale density $s(.)$. However, $\psi_{\lambda}^{\pm, \ast}(x)\neq \psi_{\lambda}^{\pm}(x), \lambda>0$ in general.
The following result gives the Laplace transform of the total occupation area.

\begin{proposition}\label{o1}
For $v_0>0$, if $S(\infty) <\infty$, then
\begin{align}
\E_{v_0} \left[ e^{-\lambda\int_0^{\infty}b^2(V_s)\1_{\left\{V_s<0\right\}}ds} \right]&=1- \frac{\int_{v_0}^{\infty}s(y)dy}{\int_0^{\infty}s(y)dy} \frac{ \psi^{+, \ast}_{\lambda}(0) }{\psi^{+, \ast}_{\lambda}(0)+\psi^{-, \ast}_{0}(0)},\label{ot1}
\end{align}
and if $S(\infty) =\infty$, then
\begin{align}
\E_{v_0} \left[ e^{-\lambda\int_0^{\infty}b^2(V_s)\1_{\left\{V_s<0\right\}}ds} \right]&=\frac{ \psi^{-, \ast}_{0}(0) }{\psi^{+, \ast}_{\lambda}(0)+\psi^{-, \ast}_{0}(0)},\label{ot2}
\end{align}

For $v_0\leq 0$,
\begin{align}
\E_{v_0} \left[ e^{-\lambda\int_0^{\infty}b^2(V_s)\1_{\left\{V_s<0\right\}}ds} \right]&=\frac{g^{\ast}_{+,\lambda}(v_0)}{g^{\ast}_{+,\lambda}(0)} \frac{ \psi^{-, \ast}_{0}(0) }{\psi^{+, \ast}_{\lambda}(0)+\psi^{-, \ast}_{0}(0)}.\label{ot3}
\end{align}
\end{proposition}
\begin{proof}
From Theorem \ref{tc} (i), we have $V_t = X_{\int_0^t b^2(V_s) ds}=X_{\varphi_t}$, $P$-a.s. on $\left\{0\leq t <\zeta  \right\}$. Since we assume that $l$ and $r$ are absorbing boundaries, thus it is understood that $\int_0^t b^2(V_s)ds=\int_0^{\zeta} b^2(V_s)ds$, $P$-a.s. on $\left\{ \infty> t\geq \zeta \right\}$. Thus we have $\int_0^{\infty} b^2(V_s)\1_{\left\{V_s<0\right\}} ds =\int_0^{\zeta} b^2(V_s)\1_{\left\{V_s<0\right\}} ds$, $P$-a.s. Apply the change of variables formula similar as Problem $3.4.5$ (vi), p174 of Karatzas and Shreve \citeyear{KS91}, we have
\begin{align}
\int_0^{\zeta} b^2(V_s)\1_{\left\{V_s<0\right\}} ds&=\int_0^{\zeta} \1_{\left\{X_{\varphi_s}<0\right\}} d\varphi_s=\int_0^{\varphi_{\zeta}} \1_{\left\{X_u<0\right\}} du=\int_0^{\zeta^X} \1_{\left\{X_u<0\right\}} du,  \text{P-a.s}
\end{align}
and the last equality is due to Theorem \ref{tc} (ii).
This combined with Lemma \ref{w1} applied to the diffusion $X$ completes the proof.  \qed
\end{proof}
\begin{remark}
Take the bankruptcy rate function as $\omega(x)=b^2(x), x<0$, then from \eqref{st}, we can obtain the \textit{probability of bankruptcy} in terms of auxiliary functions defined as combinations of solutions to the Sturm-Liouville ODE in \eqref{sl2}.
\end{remark}

\textbf{An explicit example with a general bankruptcy rate function:}

Now we show an example where we can explicitly compute the probability of bankruptcy. Assume that the company value is modeled as the following SDE with state space $J=(-\infty,\infty)$
\begin{align}
dV_t&=\mu V_t^2 dt + V_t dW_t, \quad V_0=v_0\in J,\label{mo}
\end{align}
and $\mu\neq 0$.
\begin{proposition}
For the diffusion $V$ in \eqref{mo} with the bankruptcy rate function\footnote{It is clear that $\omega(.)$ is non-negative and decreasing when $x<0$, which comply with practical applications. In our notation, we shall have $b^2(x)=\omega(x)=x^2$.} as $\omega(x)=x^2, x<0$ and $\omega(x)=0, x\geq 0$, we have that the probability of bankruptcy is given by
\begin{equation}
\psi(v_0)=
\begin{cases}
\frac{\sqrt{\mu^2+2}-\mu }{\sqrt{\mu^2+2}+\mu}e^{-2\mu v_0}, \quad &\mbox{ if } v_0>0, \mu>0\\
1-\frac{2\mu }{\sqrt{\mu^2+2}+\mu}e^{-2\mu v_0}, \quad &\mbox{ if } v_0\leq 0, \mu>0\\
1, \quad \quad \quad \quad &\mu<0
\end{cases}
\end{equation}
\end{proposition}
\begin{remark}
For the company value modeled by the diffusion $V$ in \eqref{mo}, from the above result we can see that it will eventually go bankrupt with probability $1$ if $\mu<0$. For $\mu>0$, we can explicitly determine the probability of bankruptcy.
\end{remark}
\begin{proof}
From Theorem \ref{tc}, the SDE governing the diffusion $X$ is
\begin{align}
dX_t &=\mu dt + dW_t, \quad X_0=v_0.\label{be}
\end{align}
or equivalently $X_t =v_0 + W_t +\mu t$.

From classical diffusion theory, the associated ODE \eqref{sl2} has two fundamental solutions
\begin{align}
g^{\ast}_{\pm,\lambda}(x)&=e^{\beta^{\pm}_{\lambda}}, \quad x\in \bar{J},
\end{align}
where $\beta_{\lambda}^{\pm}=-\mu\pm \sqrt{\mu^2+2\lambda}$. We can also compute
\begin{align}
S(x)&= \frac{1-e^{-2\mu x}}{2\mu}; \quad \psi_{\lambda}^{\pm, \ast}(x)=\pm \beta^{\pm}_{\lambda}.%\notag\\
%\psi_{0}^{\pm, \ast}(x)&=\pm \mu +\mid \mu\mid.
\end{align}

For $v_0>0$, if $\mu> 0$, we have $S(\infty)<\infty$. Take $\lambda=1$, then from Proposition \ref{o1}
\begin{align}
\E_{v_0} \left[ e^{-\int_0^{\infty}b^2(V_s)\1_{\left\{V_s<0\right\}}ds} \right]&=1-e^{-2\mu v_0} \frac{ \psi^{+, \ast}_{1}(0) }{\psi^{+, \ast}_{1}(0)+\psi^{-, \ast}_{0}(0)}=1-\frac{(-\mu+\sqrt{\mu^2+2})e^{-2\mu v_0} }{\mu+\sqrt{\mu^2+2}},\label{gb}
\end{align}
If $\mu\leq 0$, we have $S(\infty)=\infty$, then from Proposition \ref{o1}
\begin{align}
\E_{v_0} \left[ e^{-\int_0^{\infty}b^2(V_s)\1_{\left\{V_s<0\right\}}ds} \right]
&=\frac{ \psi^{-, \ast}_{0}(0) }{\psi^{+, \ast}_{1}(0)+\psi^{-, \ast}_{0}(0)}=0.\label{gb2}
\end{align}

For $v_0\leq 0$, if $\mu> 0$, we have $S(\infty)<\infty$ and
\begin{align}
\E_{v_0} \left[ e^{-\int_0^{\infty}b^2(V_s)\1_{\left\{V_s<0\right\}}ds} \right]
&=\frac{g^{\ast}_{+,\lambda}(v_0)}{g^{\ast}_{+,\lambda}(0)}\frac{ \psi^{-, \ast}_{0}(0) }{\psi^{+, \ast}_{1}(0)+\psi^{-, \ast}_{0}(0)}=\frac{2\mu e^{-2\mu v_0}}{\mu+\sqrt{\mu^2+2}}.\label{gb3}
\end{align}
If $\mu \leq 0$, we have $S(\infty)=\infty$ and
\begin{align}
\E_{v_0} \left[ e^{-\int_0^{\infty}b^2(V_s)\1_{\left\{V_s<0\right\}}ds} \right]
&=0.\label{gb4}
\end{align}
The above calculations combined with \eqref{st}  completes the proof. \qed
\end{proof}

\section{Stochastic drawdown area and risk model with tax }\label{s5}

%Here we apply our stochastic time change to some recent results in Zhang \citeyear{Z13}. He studies and computes the Laplace transform of the occupation time of a time-homogeneous diffusion until three kinds of stopping times: first exit time from an interval, first drawdown/drawup time, and an independent exponential time. We also have the fixed time case, which is classical, and also the case of perpetual occupation time, which is obtained by setting the parameter of exponential random variable to be $0$. Then we can deal with perpetual integral functionals with integrand weighted by its occupation indicator. This combined with the work of Salminen and Yor \citeyear{SY05} shall give us some useful results.
Assume that the stock price is modeled by a regular time-homogeneous diffusion given in \eqref{G}.
Denote $M_t =\sup_{0\leq u \leq t} V_u$ as the running maximum of the stock price. Define the first drawdown time of $a$ units as
\begin{align}
\tau_{DD}&=\inf \{t\geq 0: M_t -V_t \geq a \}. \label{dd}
\end{align}
This stopping time is of importance in modeling stock prices during downturn of the financial market(e.g. the 2008 financial crisis), and is an inherent constraint in some portfolio optimization problems. The drawdown constraint allows us to encode risk attitudes into the portfolio optimization problem, and is of both practical and theoretical interests. It was first introduced by Grossman and Zhou \citeyear{GZ93} in a continuous-time framework , and studied by Cvitanic and Karatzas \citeyear{CK95} and Cherny and Obloj \citeyear{CO13}.

The seminal paper Lehoczky \citeyear{L77} provides a closed-form expression for the joint Laplace transform of the first drawdown time and the running maximum stopped at the first drawdown time.
\begin{lemma}\label{leho}(equation $(4)$, p$602$ of Lehoczky \citeyear{L77})

The joint Laplace transform of $\tau_{DD}$ and $M_{\tau_{DD}}$ is
\begin{align}
E[e^{-\alpha M_{\tau_{DD}} -\beta \tau_{DD}}]&=\int_0^{\infty} e^{-\alpha u -\int_0^u d(z) dz} c(u) du.\label{le}
\end{align}
for $\alpha, \beta\geq 0$, where
\begin{align}
d(z)&=\frac{g(z-a)h^{\prime}(z)-h(z-a)g^{\prime}(z)}{g(z-a)h(z)-g(z)h(z-a)};\notag\\
c(x)&=\frac{g(x)h^{\prime}(x)-g^{\prime}(x)h(x)}{g(x-a)h(x)-g(x)h(x-a)}.
\end{align}
and here $g(.)$ and $h(.)$ are two independent solutions to the following Sturm-Liouville ODE associated with diffusion $V$.
\begin{align}
\frac{1}{2}\sigma^2(x) f^{\prime\prime}(x)+ \mu(x) f^{\prime}(x) &=\beta f(x),\quad  x\in[-a,\infty).
\end{align}
\end{lemma}

Denote the stochastic area till the first drawdown time of $a$ units as $\int_0^{\tau_{DD}} b^2(V_s)ds$.
Similar as above, we can derive the joint Laplace transform of $M_{\tau_{DD}}$ and $\int_0^{\tau_{DD}} b^2(V_s)ds$.
\begin{proposition}
The joint Laplace transform of $M_{\tau_{DD}}$ and $\int_0^{\tau_{DD}} b^2(V_s)ds$ is
\begin{align}
E[e^{-\alpha M_{\tau_{DD}} -\beta \int_0^{\tau_{DD}} b^2(V_s)ds}]&=\int_0^{\infty} e^{-\alpha x -\int_0^x d^{\ast}(z) dz} c^{\ast}(x) dx.\label{le2}
\end{align}
for $\alpha, \beta\geq 0$, where
\begin{align}
d^{\ast}(z)&=\frac{g^{\ast}(z-a)h^{\prime \ast}(z)-h^{\ast}(z-a)g^{\prime \ast}(z)}{g^{\ast}(z-a)h^{\ast}(z)-g^{\ast}(z)h^{\ast}(z-a)};\notag\\
c^{\ast}(x)&=\frac{g^{\ast}(x)h^{\prime \ast}(x)-g^{\prime \ast}(x)h^{\ast}(x)}{g^{\ast}(x-a)h^{\ast}(x)-g^{\ast}(x)h^{\ast}(x-a)}.
\end{align}
and here $g^{\ast}(.)$ and $h^{\ast}(.)$ are any two independent solutions to the following ODE
\begin{align}
\frac{1}{2}\frac{\sigma^2(x)}{b^2(x)} f^{\prime\prime}(x)+\frac{\mu(x)}{b^2(x)} f^{\prime}(x)  &=\beta  f(x), \quad x\in[-a,\infty).
\end{align}
\end{proposition}
\begin{proof}
From Theorem \ref{tc} (i), we have $V_t = X_{\int_0^t b^2(V_s) ds}$, $P$-a.s. on $\left\{0\leq t <\zeta  \right\}$. Define $\tau^{X}_{DD}=\inf \{t\geq 0: \max_{0\leq u \leq t} X_t -X_t \geq a \}$, then from Theorem \ref{tc} (iii), we have $\tau^{X}_{DD}=\int_0^{\tau_{DD}} b^2(V_s)ds$, $P$-a.s., and $\max_{0\leq u\leq \tau_{DD}} V_u =\max_{0\leq u\leq \tau^X_{DD}} X_u$, $P$-a.s. This combined with Lemma \ref{leho} completes the proof.  \qed
\end{proof}

%\subsection{Stochastic area till the Azema-Yor stopping time and diffusion risk model with tax}
Azema and Yor \citeyear{AY79} introduced a family of simple local martingales and proposed a solution to the Skorokhod embedding problem. These processes are later named Azema-Yor processes and the associated passage time is called the Azema-Yor stopping time. Their applications range from solving the Skorokhod embedding problem(Obloj \citeyear{O04}), pricing capped Russian options(Ott \citeyear{O13}) and portfolio optimization with drawdown constraints(El Karoui and Meziou \citeyear{EM06})

%Consider a regular time-homogeneous diffusion $V$ in \eqref{G}, and denote $M_t:=\max_{0\leq u\leq t}V_u$ its running maximum. The Azema-Yor stopping time is
Consider the diffusion $V$ as defined in \eqref{G} with initial value $v_0>0$, and define the running maximum process associated with $V$ as
\begin{align}
M_t &=(\max_{0\leq u \leq t} V_u)\vee s.\label{rm}
\end{align}
started at $s\geq v_0 >0$. Define the Azema-Yor stopping time as
\begin{align}
\tau_{AY} &=\inf \{t>0:  V_t \leq g(M_t)  \},
\end{align}
for any continuous function $g$ defined on $[0,\infty)$ satisfying $0<g(x)<x$ for $x>0$.

In the following we show another application of the Azema-Yor stopping time in a diffusion risk model where we assume that there is a loss-carry-forward taxation.  It was introduced into the risk theory by Albrecher and Hipp \citeyear{AH07} in the Levy insurance model framework. The basic idea is to allow the tax to be paid at a certain fixed rate \textit{immediately} when the surplus of the company is at a running maximum. For a pointer to the literature in this area, please refer to Albrecher, Renaud and Zhou \citeyear{ARZ08} and the references therein.

We cast our model in a regular time-homogeneous diffusion setting(Li, Tang and Zhou \citeyear{LTZ13}).
Assume that the value of the firm is modeled by the diffusion $V$ in \eqref{G}. Now introduce a surplus-dependent tax rate: whenever the process $V_t$ coincides with its running maximum $M_t$, the firm pays tax at rate $\gamma(M_t)$ and $\gamma(.):[v_0, \infty)\rightarrow [0,1)$ is a measurable function. The value process after taxation satisfies
\begin{align}
dU_t &=dV_t -\gamma(M_t) dM_t, \quad  t\geq 0,\label{U}
\end{align}
with $U_0 =V_0 =v_0$.  For a default threshold $a$(conventionally assigned $0$ in the ruin theory), define the \textit{time of default with tax} as
\begin{align}
T^U (a) &=\inf\{t\geq 0 : U_t = a\},\label{fptu}
\end{align}
and $\inf \emptyset =\infty$ by convention. Note that $a< v_0$. Now we want to compute $\E[T^U (a)]$, which represents the expected time of ruin. Introduce the following function
\begin{align}
\bar{\gamma}(x)&=x-\int_{v_0}^x \gamma(z)dz=v_0 +\int_{v_0}^x (1-\gamma(z))dz, \quad x\geq v_0 .
\end{align}
Notice that $v_0 < \bar{\gamma}(x)\leq x$. We have the following representation $U_t =V_t -M_t +\bar{\gamma}(M_t)$, then we have
\begin{align}
T^U (a) &=\inf\{t\geq 0 : U_t = a\}=\inf\{t\geq 0, V_t = g(M_t) \},\label{t1}
\end{align}
where
\begin{align}
g(x)=x-\bar{\gamma}(x)+a=\int_{v_0}^x \gamma(z)dz+a.\label{ge}
\end{align}
We have that $a\leq g(x)<x-v_0+a<x$ because $\gamma(.):[v_0, \infty)\rightarrow [0,1)$. Thus we can see that equation \eqref{t1} represents an Azema-Yor stopping time.

Our objective is to calculate the expectation of the \textit{expected time of ruin} $T^U(a)$ and the \textit{stochastic area till ruin} $\int_0^{T^U(a)} b^2(V_s)ds$.
 We first recall a lemma.
\begin{lemma}\label{pp}(Theorem $4.1$ of Pedersen and Peskir \citeyear{PP98})
Here $s\geq v_0$ is the initial value of the running maximum process $M$.
If $g(s)<v_0\leq s$, then
\begin{align}
\E[\tau_{AY}]&=2\frac{S(s)-S(v_0)}{S(s)-S(g(s))}\int_{g(s)}^{v_0} \frac{S(t)-S(g(s))}{\sigma^2 (t) s(t)} dt +2\frac{S(v_0)-S(g(s))}{S(s)-S(g(s))} \left\{\int_{v_0}^{s} \frac{S(s)-S(t)}{\sigma^2(t)s(t)}dt \right.\notag\\
&\left.+\int_{s}^{\infty} \frac{s(t)}{S(t)-S(g(t))}\( \int_{g(t)}^t \frac{S(r)-S(g(t))}{\sigma^2(r)s(r)}dr   \)\exp\( -\int_{s}^t \frac{s(r)}{S(r)-S(g(r))}dr  \)dt   \right\}.\label{pp2}
\end{align}
and $\E[\tau_{AY}]=0$ for $0<v_0\leq g(s)$.
\end{lemma}
For the diffusion risk model with tax, we have the following result on the expected value of the ruin time and the stochastic area till ruin.

%\begin{proposition}
%If $g(v_0)<x\leq v_0$, then
%\begin{align}
%&E[\int_0^{\tau_{AY}} b^2(V_s)ds ]\notag\\
%&=2\frac{S(v_0)-S(x)}{S(v_0)-S(g(v_0))}\int_{g(v_0)}^{x} \frac{b^2(t)(S(t)-S(g(v_0)))}{\sigma^2 (t) s(t)} dt \notag\\
%&+2\frac{S(x)-S(g(v_0))}{S(v_0)-S(g(v_0))} \(\int_{x}^{v_0} \frac{b^2(t)(S(v_0)-S(t))}{\sigma^2(t)s(t)}dt \right.\notag\\
%&\left.+\int_{v_0}^{\infty} \frac{s(t)}{S(t)-S(g(t))}\( \int_{g(t)}^t \frac{b^2(r)(S(r)-S(g(t)))}{\sigma^2(r)s(r)}dr   \)\exp\( -\int_{v_0}^t \frac{s(r)}{S(r)-S(g(r))}dr  \)dt   \),
%\end{align}
%and $E[\int_0^{\tau_{AY}} b^2(V_s)ds ]=0$, for $0<x\leq g(v_0)$.
%\end{proposition}
%\begin{remark}
%The above proposition characterizes the first moment of the integral functional of the diffusion $V$ stopped at the Azema-Yor stopping time $\tau_g$. It is of interest when we want to compute the optimized value function for some optimal stopping problems. And in these problems, the optimal stopping time is usually the Azema-Yor stopping time as defined before.
%\end{remark}
\begin{proposition}(Expected Ruin Time and Ruin Area with Tax)

With $g(.)$ defined in \eqref{ge}

(i) The expected time of ruin with tax is
\begin{align}
\E[T^U (a)]&=2\int_{v_0}^{\infty} \frac{s(t)}{S(t)-S(g(t))}\( \int_{g(t)}^t \frac{S(r)-S(g(t))}{\sigma^2(r)s(r)}dr \)\exp\( -\int_{v_0}^t \frac{s(r)}{S(r)-S(g(r))}dr  \)dt .\label{pp3}
\end{align}
(ii) The expected stochastic area till the ruin time with tax is
\begin{align}
&\E\left[\int_0^{T^U (a)} b^2(V_s)ds \right]\notag\\
&=2\int_{v_0}^{\infty} \frac{s(t)}{S(t)-S(g(t))}\( \int_{g(t)}^t \frac{b^2(r)(S(r)-S(g(t)))}{\sigma^2(r)s(r)}dr   \)\exp\( -\int_{v_0}^t \frac{s(r)}{S(r)-S(g(r))}dr  \)dt .\label{pp4}
\end{align}
\end{proposition}
\begin{proof}
For (i), note that $U_0=v_0$ means that we consider the case when the running maximum starts at the same initial value of the process $V$, i.e. $s=v_0$ in \eqref{rm}. Taking $s=v_0$ in \eqref{pp2}, we arrive at the expression \eqref{pp3}.

For (ii), from Theorem \ref{tc} (i), we have $V_t = X_{\int_0^t b^2(V_s) ds}$, $P$-a.s. on $\left\{0\leq t <\zeta  \right\}$. Define $T_X^U (a)=\inf\{t\geq 0, X_t \leq g(M^X_t) \}$, then from Theorem \ref{tc} (iii), we have $T_X^U (a)=\int_0^{T^U (a)} b^2(V_s)ds$, $P$-a.s., and $\max_{0\leq u\leq T^U (a)} V_u =\max_{0\leq u\leq T_X^U (a)} X_u$, $P$-a.s. This combined with part (i) completes the proof.  \qed
\end{proof}

%\textbf{Example: geometric Brownian motion with surplus-dependent tax}
%
%
%Assume that the firm value process is modeled by a geometric Brownian motion in \eqref{gbms} with $\mu\neq 0$, and assume that $\nu=\frac{2\mu}{\sigma^2}-1>0$. The scale density and scale function are $s(x)= c_3 x^{-\nu-1}, S(x)=-c_3 \frac{x^{-\nu}}{\nu}$, where $c_3$ is a constant. Choose $b(x)=x^2$, and consider a constant tax rate $\gamma(z)=\gamma\in[0,1)$. Then $g(x)=\gamma (x-v_0)+a, x\in[v_0, \infty)$,  and we have
%\begin{align}
%\E[T^U (a)]&=2\int_{v_0}^{\infty} \frac{s(t)}{S(t)-S(g(t))}\( \int_{g(t)}^t \frac{S(r)-S(g(t))}{\sigma^2(r)s(r)}dr \)\exp\( -\int_{v_0}^t \frac{s(r)}{S(r)-S(g(r))}dr  \)dt \notag\\
%&=
%\end{align}

\section{Conclusion and future research}\label{s6}

In this paper we have studied the stochastic area of a time-homogeneous diffusion till a stopping time. Through stochastic time change we explicitly express the Laplace transform of the stochastic area in terms of eigenfunctions of an associated diffusion. We also obtain the integer moments of the stochastic area explicitly in terms of scale and speed densities, and generalize the work of Abundo \citeyear{A133} \citeyear{A13} in the case of time-homogeneous diffusions, since his approach requires solving a partial differential-difference equation with outer conditions. We generalize the work of Gerber, Shiu and Yang \citeyear{GSY12} by computing the ruin probability for the Omega risk model with a general bankruptcy rate function and we also obtain an explicit expression for the expected time of ruin in a diffusion risk model with tax. Future research is to extend the study to incorporate jump diffusions, such as the mixed exponential jump diffusion model proposed in Cai and Kou \citeyear{CK11}.

\bibliographystyle{spbasic}

\bibliography{area}

\end{document}